\documentclass[9pt, twocolumn, fleqn]{article}

\usepackage[a4paper, margin=1.5cm]{geometry}

\usepackage{latexsym}
\usepackage{amsmath}
\usepackage{amsfonts}
\usepackage{amssymb}
\usepackage{graphicx}
\usepackage{slashbox}
\usepackage{color}
\usepackage{tikz}
\usepgflibrary{arrows}
\usepackage{mathbbol}
\let\OLDthebibliography\thebibliography
\renewcommand\thebibliography[1]{
  \OLDthebibliography{#1}
  \setlength{\parskip}{0pt}
  \setlength{\itemsep}{0pt}
}

\title{Conformal symmetry, chiral fermions \\and semiclassical approximation}
\author{
Joanna Gonera\footnote{joanna.gonera@uni.lodz.pl}, Piotr Kosi\'nski\footnote {piotr.kosinski@uni.lodz.pl}, Pawe\l{} Ma\'slanka\footnote{pawel.maslanka@uni.lodz.pl}\\
\small \textit{Department of Computer Science} \\
\small \textit {Faculty of Physics and Applied Informatics}\\
\small \textit {University of Lodz, Poland}\\
}

\date{}
\begin{document}
\maketitle
\begin{abstract}
The explicit form of conformal generators is found which provides the extension of Poincare symmetry for massless particles of arbitrary helicity. The helicity $\frac {1}{2}$ particles are considered as the particular example. The realization of conformal symmetry in the semiclassical regime of Weyl equation is obtained.
\end{abstract}

\section{Introduction} 
\label{I}
\par In recent years the dynamics of chiral fermions became a field of intense research \cite{b1}-\cite{b20}. This topic is interesting due to its applications, for example in exploring anomaly-related phenomena in kinetic theory, but also because its complete theoretical description requires to combine the diverse ideas concerning topology, symmetries, anomalies, Berry's phases etc.
\par In the most important case of helicity $\pm\frac{1}{2}$ massless fermions the starting point is the Weyl equation in electromagnetic field; the latter is typically viewed as classical external one, at least in the context we are concerned here. For many applications the electromagnetic field is assumed to be weak and slowly varying which allows for the use of semiclassical approximation. The semiclassical dynamics can be derived either from path-integral representation of transition amplitudes \cite{b11} or using time-dependent variational principle \cite{b20} popular among condensed matter physicists \cite{b21}-\cite{b23}. The result is summarized in semiclassical action functional involving canonical variables describing chiral fermion (see eqs. (\ref{al29}), (\ref{al30}) below).
\par One of the important problems inherent in semiclassical approach is that of Poincare covariance. Weyl equation enjoys standard Poincare covariance. However, the semiclassical action functional lacks manifest Lorentz covariance (which becomes Lorentz symmetry in the absence of external fields), in spite of the fact that it is derived from explicitly covariant Weyl equation. The way to cure this situation was suggested in Ref. \cite{b11} where a modified transformation rule under Lorentz boosts has been proposed. It involves $ O(\hslash) $ corrections which are helicity-dependent and provide on-shell implementation of Poincare covariance on semiclassical level. The resulting situation attracted much attention \cite{b9}-\cite{b11}, \cite{b16}-\cite{b18}, \cite{b19}, \cite{b20}. Duval, Horvathy et al. \cite{b16}-\cite{b18} derived the modified transformation rules from Souriau model \cite{b24} of spinning particle using the procedure called spin enslaving. Similar results were obtained by using coadjoint orbit method \cite{b19}. On the other hand, the time-dependent variational principle provides particularly clear interpretation of canonical variables entering semiclassical action which allows for derivation of the relevant transformation rules from first principles \cite{b20}.
\par It is well known (see, for example, Ref. \cite{b25}) that the massless representations of Poincare group can be extended to those of the conformal one. On the helicity $ \pm\frac {1}{2} $ level this is equivalent to the conformal covariance of Weyl equation (which persists in the presence of external electromagnetic field) \cite{b26}.
\par The natural question arises if the conformal symmetry survives the semiclassical approximation and what is the actual form of relevant transformations. 
\par The present paper is addressed to this problem. In Sec. II we present a simple method of constructing classical and quantum conformally invariant Hamiltonian dynamics of massless particles. The particular example of helicity $ -\frac {1}{2} $ fermion is discussed in Sec. III. The results obtained are then used in Sec. IV to find semiclassical form of conformal symmetry. Some conclusions are presented in final Sec. V.
\section{Conformally invariant dynamics}
\label{II}
\par It is well known that the Hamiltonian systems admitting transitive action of some symmetry group can be classified using coadjoint orbit method \cite{b24}, \cite{b29}. The resulting Hamiltonian dynamics may be then quantized leading in many cases to consistent quantum theory. The procedure outlined is quite effective. Given a symmetry group $ G $ it yields, in principle, all "elementary" systems, classical and quantum, enjoying the symmetry described by $ G $. 
\par The key ingredient of the orbit method is the classification of all coadjoint orbits. The generic ones are obtained as invariant submanifolds in the space dual to relevant Lie algebra, corresponding to the fixed values of Casimir functions. However, as we shall see below, the nongeneric orbits are also relevant from the physical point of view. Its effective characterization is slightly more involved. They are defined by the sets of equations invariant under coadjoint action of the group. The infinitesimal coadjoint action is defined in terms of the relevant Poisson brackets. Therefore, the set of equations defining a given orbit must be invariant under the operation of taking Poisson bracket with any coordinate function on the dual space. In other words, the corresponding set of functions generates the ideal $ I $ in the algebra of functions on dual space obeying 
\begin {align} 
\label {al1}
\{\cdot, I\}\subset I 
\end {align}
where the dot stands for any element of the algebra of functions.
\par The importance of nongeneric orbits is clearly seen on the example of Poincare group. The generic orbits are here eightdimensional. However, the orbits corresponding to massless particles are sixdimensional because there are no spin degrees of freedom. Moreover, if the Casimir function related to the mass squared vanishes then the second Casimir also vanishes; consequently, helicity is not characterized by the values of Casimirs. 
\par Sixdimensional orbits are nongeneric; we have to impose four constraints in order to define the corresponding submanifolds. To this end let \linebreak $ \zeta_{\mu}(\zeta_{\mu\nu}=-\zeta_{\nu\mu}) $ be the coordinates in the space dual to Poincare algebra, corresponding to the generators $ P_{\mu}(M_{\mu\nu}=-M_{\nu\mu}) $. The relevant Poisson brackets read:
\begin {align} 
\label {al2}
\{\zeta_{\mu},\zeta_{\nu}\}&=0 \nonumber\\
\{\zeta_{\mu\nu},\zeta_{\rho}\}&=g_{\nu\rho}\zeta_{\mu}-g_{\mu\rho}\zeta_{\nu}\nonumber\\
\{\zeta_{\mu\nu}, \zeta_{\rho\sigma}\}&=g_{\mu\sigma}\zeta_{\nu\rho}+g_{\nu\rho}\zeta_{\mu\sigma}-g_{\mu\rho}\zeta_{\nu\sigma}-g_{\nu\sigma}\zeta_{\mu\rho}
\end {align}
The ideals defining the relevant orbits are generated by 
\begin {align} 
\label {al3}
J_{1}\equiv &\zeta_{0}^{2}-\zeta_{k}^{2}\quad \text {(summation over $k$ assumed)}\nonumber\\
J_{2(ij)}\equiv &-J_{2(ji)}\equiv\zeta_{ij}-\frac{\zeta_{i}\zeta_{0j}}{\zeta_{0}}+\frac{\zeta_{j}\zeta_{0i}}{\zeta_{0}}+\lambda\epsilon_{ijk}\dfrac{\zeta_{k}}{\zeta_{0}}
\end {align}
\par It is easy to verify that $J_{1}$, $J_{2(ij)}$, $i,j=1,2,3$, generate the ideal obeying (\ref{al1}); $J_{1}$ is simply the first Casimir function while $J_{2(ij)}$ depend on one real parameter $\lambda$, the classical helicity. One can choose $\zeta_{0i}$ and $\zeta _{i}$, $i=1,2,3$, as independent coordinates on the orbit; then
\begin {align} 
\label {al4}
&\zeta_{0}=\pm(\zeta_{k}\zeta_{k})^{\frac{1}{2}}\nonumber\\
&\zeta_{ij}=\frac{\zeta_{i}\zeta_{0j}}{\zeta_{0}}-\frac{\zeta_{j}\zeta_{0i}}{\zeta_{0}}-\lambda\epsilon_{ijk}\frac{\zeta_{k}}{\zeta_{0}}
\end {align}
and, as in the standard approach, one can show that the choice of sign in the first eq. (\ref{al4}) corresponds to the choice of connected component of the orbit. 
\par Alternatively, one may use the almost Darboux coordinates defined as
\begin {align} 
\label {al5}
&p_{i}\equiv\zeta_{i}\nonumber\\
&x^{i}\equiv \frac{\zeta_{0i}}{\zeta_{0}}
\end {align}
Then eqs. (\ref{al2}) imply
\begin {align} 
\label {al6}
&\{p_{i},p_{j}\}=0\nonumber\\
&\{x^{i},p_{j}\}=-\delta_{ij}
\end {align}
\par In order to find the remaining Poisson bracket we use again \linebreak eqs. (\ref{al2}) together with eqs. (\ref{al4}); the result reads
\begin {align} 
\label {al7}
\{x^{i},x^{j}\}=\lambda\epsilon_{ijk}\frac{p_{k}}{\vert\vec{p}\vert^{3}}
\end {align}
\par The Lorentz generators $M_{\mu\nu}(\equiv \zeta_{\mu\nu})$, when expressed in terms of new coordinates, read
\begin {align} 
\label {al8}
&M_{ij}=p_{i}x^{j}-p_{j}x^{i}-\lambda\epsilon_{ijk}\frac{p_{k}}{\vert\vec{p}\vert}\nonumber\\
&M_{0i}=\vert\vec{p}\vert x^{i}
\end {align}
One easily verifies that the Lorentz algebra is obeyed. \linebreak
Let us now quantize the resulting Poisson algebra. To this end we put
\begin {align} 
\label {al9}
(\hat{p}_{i}\phi)(p)=p_{i}\phi(p)
\end {align}
and the scalar product is defined simply as 
\begin {align} 
\label {al10}
(\phi,\psi)=\int d^{3}\vec{p}\,\overline{\phi(p)}\psi(p)
\end {align}
Then
\begin {align} 
\label {al11}
\hat{x}^{i}=i\frac{\partial}{\partial p_{i}}+A_{i}(p)\equiv iD_{i}
\end {align}
where $A_{i}(p)$ is a vector potential for monopole of charge $\lambda$.
\par In order to find the Lorentz generators it is sufficient to write out the properly ordered version of eqs. (\ref{al8}). It appears that no ordering is necessary for rotation generators while the simplest ordering leading to hermitean expressions works for boosts:
\begin {align} 
\label {al12}
&\hat {M}_{ij}=p_{i}(iD_{j})-p_{j}(iD_{i})-\lambda\epsilon_{ijk}\frac{p_{k}}{\vert\vec{p}\vert}\nonumber\\
&\hat {M} _{0k}=\vert \vec{p}\vert (iD_{k})-\frac{ip_{k}}{2\vert\vec{p}\vert}
\end {align}
Eqs. (\ref{al12}) agree with the standard expressions \cite{b27}, \cite{b28}.
\par We would like to extend our representation to that for conformal algebra. The dual space is now fifteendimensional so we need five additional generators to define the relevant ideal. Let $\eta$ be the coordinate corresponding to dilatation generator $D$ while       $\eta_{\mu}$ represent the generators $K_{\mu}$ of special conformal transformations. The Poisson brackets (\ref{al2}) must be supplemented by 
\begin {align} 
\label {al13}
\{\eta,\zeta_{\mu}\}&=\zeta_{\mu}\nonumber\\
\{\eta,\zeta_{\mu\nu}\}&=0\nonumber\\
\{\eta,\eta_{\mu}\}&=-\eta_{\mu}\nonumber\\
\{\zeta_{\mu\nu},\eta_{\rho}\}&=g_{\rho\nu}\eta_{\mu}-g_{\rho\mu}\eta_{\nu}\nonumber\\
\{\eta_{\mu}, \zeta_{\nu}\}&=2(\zeta_{\mu\nu}-g_{\mu\nu}\eta)\nonumber\\
\{\eta_{\mu},\eta_{\nu}\}&=0
\end {align}
It is not difficult to verify that the following expressions generate, together with $J_{1}$, $J_{2(ij)}$, the invariant ideal:
\begin {align} 
\label {al14}
J_{3}\equiv &\eta+\frac{\zeta_{k}\zeta_{0k}}{\zeta_{0}}\nonumber\\
J_{4(0)}\equiv &\eta_{0}+\frac{\zeta_{0k}\zeta_{0k}}{\zeta_{0}}+\frac{\lambda^{2}}{\zeta_{0}}\nonumber\\
J_{4(i)}\equiv&\eta_{i}+\frac{\zeta_{i}\zeta_{0k}\zeta_{0k}}{\zeta_{0}^{\phantom {0}2}}-2\frac{\zeta_{0i}\zeta_{k}\zeta_{0k}}{\zeta_{0}^{\phantom {0}2}}-2\lambda\epsilon_{ikl}\frac{\zeta_{0k}\zeta_{l}}{\zeta_{0}^{\phantom {0}2}}-\lambda^{2}\frac{\zeta_{i}}{\zeta_{0}^{\phantom {0}2}}
\end {align}
Correspondingly, the additional generators, expressed in terms of $p_{i}$, $x^{i}$, read:
\begin {align} 
\label {al15}
D=&-p_{k}x^{k}\nonumber\\
K_{0}=&-\vert\vec{p}\vert x^{k}x^{k}-\frac{\lambda^{2}}{\vert\vec{p}\vert}\nonumber\\
K_{i}=&-p_{i}x^{k}x^{k}+2p_{k}x^{k}x^{i}+2\lambda\epsilon_{ikl}\frac{x^{k}p_{l}}{\vert\vec{p}\vert}+\frac{\lambda^{2}p_{i}}{\vert\vec{p}\vert^{2}}
\end {align}
By virtue of eqs. (\ref{al6}) and (\ref{al7}), the generators (\ref{al8}), (\ref{al15}) span the Lie algebra of conformal group (with Lie bracket being the Poisson bracket).
\par The next step is to find the quantum counterpart of eqs. (\ref{al15}). To this end we replace $x^{i}$ and $p_{i}$ by the relevant operators and look for the proper ordering on the right hand sides of eqs. (\ref{al15}). This is slightly more troublesome than in the case of Poincare generators (\ref{al12}); in particular, the terms proportional to $\lambda^{2}$ acquire quantum corrections. We obtain 
\begin {align} 
\label {al16}
\hat{D}=&-p_{k}(iD_{k})+\frac{3i}{2}\nonumber\\
\hat{K}_{0}=&-(iD_{k})\vert\vec{p}\vert(iD_{k})+\bigg(\frac{3}{4}-\lambda^{2}\bigg)\frac{1}{\vert\vec{p}\vert}\nonumber\\
\hat{K}_{i}=&-(iD_{k})p_{i}(iD_{k})+(iD_{k})p_{k}(iD_{i})+(iD_{i})p_{k}(iD_{k})+\nonumber\\
&+2\lambda\epsilon_{ikl}\frac{p_{l}}{\vert\vec{p}\vert}(iD_{k})+\bigg(\lambda^{2}+\frac{1}{4}\bigg)\frac{p_{i}}{\vert\vec{p}\vert^{2}}
\end {align}
\par The painstaking computations show that the generators (\ref{al8}) and (\ref{al16}) obey the commutation rules of conformal algebra. In particular, eqs. (\ref{al16}) show how one can accommodate the representation of conformal algebra in the representation space of Poincare covariant massless helicity $\lambda$ particles.
\par Finally, let us note that the classical limit (\ref{al15}) is obtained from (\ref{al16}) by restoring the explicit $\hslash$-dependence, identifying $i\hslash D_{k}$ with $\hat{x}^{k}$ and taking $\hslash\rightarrow 0$, $\lambda\rightarrow\infty$, $\hslash x$ - fixed $(\equiv\lambda)$.
\section{The $\vert \lambda \vert = \frac {1}{2}$ case} 
\label{III}
\par Let us now consider the helicity $\vert \lambda\vert = \frac{1}{2}$ fermions. They play distinguished role both because of their physical relevance and elegance of mathematical description. The (say) left-handed fermions are described by the Weyl equation
\textsc{\begin {align} 
\label {al17}
\sigma^{\mu}\partial_{\mu}\psi=0
\end {align}}
with $\sigma^{\mu}=(\mathbb{1},-\vec{\sigma})$, $\vec{\sigma}=(\sigma_{1},\sigma_{2}, \sigma_{3})$ being Pauli matrices. The single-particle theory is based on positive energy solutions to eq. (\ref{al17}):
\begin {align} 
\label {al18}
\psi (\vec{x},t)=\frac{1}{(2\pi)^{3/2}}\int d^{3}\vec{p}u_{+}(p)c(p)e^{i(\vec{p}\cdot \vec{x}-\vert \vec{p}\vert t)}
\end {align}
with $u_{+}(p)$ being the positive energy spinor obeying
\begin {align} 
\label {al19}
&\vec{p}\cdot \vec{\sigma}u_{+}(p)=-\vert \vec{p}\vert u_{+}(\vec{p}), \quad u^{+}_{+}(p)u_{+}(p)=1;
\end {align}
in particular, one can choose
\begin {align} 
\label {al20}
u_{+}(p)=\frac{1}{\sqrt{2\vert\vec{p}\vert (\vert \vec{p}\vert +p^{3})}} \binom {p^{1}-ip^{2}} {-(\vert \vec {p}\vert +p^{3})}
\end {align}
\par Weyl equation (\ref{al17}) is purely kinematical in the sense that it is equivalent to the statement that $c(p)$ span massless $\lambda = -\frac{1}{2}$ unitary representation of Poincare group. However, its advantage is that it admits straightforward coupling to electromagnetic field in covariant and casual way. On the other hand, for general helicity $\lambda$ the relevant wave equation (carrying the information about the representation of Poincare group) is of order $2 \vert \lambda \vert $ \cite{b30} which makes the correct coupling to electromagnetism more problematic.
\par It is well known that the Weyl  equation (\ref{al17}) is invariant under the space-time action of conformal group generated by
\begin {align} 
\label {al21}
P_{\mu}=&i\partial_{\mu}\nonumber\\
M_{\mu\nu}=&i(x_{\mu}\partial_{\nu}-x_{\nu}\partial _{\mu})+\Sigma _{\mu\nu}, \nonumber\\
&\Sigma_{0k}=\frac{i \sigma_{k}}{2}, \quad \Sigma_{ij}=\frac{1}{2}\epsilon_{ijk}\sigma_{k}\nonumber\\
D=&-ix^{\mu}\partial_{\mu}-\frac{3i}{2}\nonumber\\
K_{\mu}=&i(x^{2}\partial_{\mu}-2x_{\mu}x^{\nu}\partial_{\nu})-3ix_{\mu}-2x^{\nu}\Sigma_{\mu\nu}
\end {align}
\par The action of the generators (\ref{al21}) on the wave function (\ref{al18}) may be expressed as the action of generators (\ref{al12}) and (\ref{al16}) on $c(p)$, with $\lambda=-\frac{1}{2}$ and
\begin {align} 
\label {al22}
iD_{k}=i\frac{\partial}{\partial p^{k}}+iu^{+}_{+}(p)\frac{\partial u_{+}(p)}{\partial p^{k}}
\end {align}
\par The Weyl equation (\ref{al17}) is derived from the action functional
\begin {align} 
\label {al23}
S=i\int d^{4}x \psi^{+}\sigma^{\mu}\partial_{\mu}\psi
\end {align}
By considering the general form of variation of $S$ one finds the conserved charge following from its invariance
\begin {align} 
\label {al24}
Q=i  \int  d^{3}\vec{x}\psi^{+}(\delta\psi-\delta x^{k}\partial_{k}\psi+\delta x^{0}\sigma^{k}\partial_{k}\psi)
\end {align}
with $\delta\psi=\delta_{0}\psi+\delta x^{\mu}\partial_{\mu}\psi$.\newline
The off-shell form of (\ref{al18}) reads
\begin {align} 
\label {al25}
\psi (x)=\frac{1}{(2\pi)^{3/2}}\int d^{3}\vec{p}u_{+}(p)c(\vec{p},t)e^{i\vec{p}\cdot\vec{x}}
\end {align}
and the Weyl equation is equivalent to $\dot{c}(\vec{p},t)=-i\vert\vec{p}\vert c (\vec{p},t)$.
\par Inserting $\delta_{0}\psi\sim G\psi$ with $G=P_{\mu},M_{\mu\nu},D,K_{\mu}$ \big(cf. eqs. (\ref{al21})\big) into  eq. (\ref{al24}) we find the conserved charges following from conformal symmetry
\begin {align} 
\label {al26}
\mathcal{P}_{k}=&\int  d^{3}\vec{p}p_{k}\vert c(\vec{p},t)\vert^{2}\nonumber\\
\mathcal{P}_{0}=&\int  d^{3}\vec{p}\vert \vec{p}\vert \vert c(\vec{p},t)\vert^{2}\nonumber\\
\mathcal{M}_{ij}=&\int  d^{3}\vec{p\,}\overline{c(\vec{p},t)} \bigg ( p_{i}(iD_{j})-p_{j}(iD_{i})+\frac{\epsilon_{ijk}p_{k}}{2\vert\vec{p}\vert}\bigg )c(\vec{p},t)\nonumber\\
\mathcal{M}_{0i}= &\int  d^{3}\vec{p}\,\overline{c(\vec{p},t)}\bigg (\vert \vec{p}\vert (iD_{i})-\frac{ip_{i}}{2\vert \vec{p}\vert}\bigg)c (\vec{p},t)+\nonumber\\
&+x^{0}\int  d^{3}\vec{p}p_{i}\vert c(\vec{p},t)\vert ^{2}\nonumber\\
\mathcal{D}=&\int  d^{3}\vec{p}\,\overline{c(\vec{p},t)}\bigg(\!\!-p_{k}(iD_{k})+\frac{3i}{2}\bigg) c(\vec{p},t)+\nonumber\\
&-x^{0}\int d^{3}\vec{p}\vert \vec{p}\vert \vert c(\vec{p},t)\vert^{2}\nonumber\\
\mathcal{K}_{0}=&\int d^{3}\vec{p}\,\overline{c(\vec{p},t)} \bigg(\!\!-(iD_{k})\vert\vec{p}\vert (iD_{k})+\frac{1}{2\vert\vec{p}\vert}\bigg) c(\vec{p},t)+\nonumber \\
&+ 2x^{0}\int d^{3}\vec{p}\,\overline{c(\vec{p},t)}\bigg (\!\!-p_{k}(iD_{k})+\frac{3i}{2}\bigg ) c(\vec{p},t)+\nonumber\\
&-(x^{0})^{2}\int d^{3} \vec{p}\vert\vec{p}\vert \vert c(\vec{p},t)\vert^{2}\nonumber\\
\mathcal{K}_{i}=&\int d^{3}\vec{p}\,\overline{c(\vec{p},t)}\bigg(\!\!-(iD_{k})p_{i}(iD_{k})+\nonumber\\
&+(iD_{k})p_{k}(iD_{i})+(iD_{i})p_{k}(iD_{k})+\nonumber\\
&-\epsilon_{ikl}\frac{p_{l}}{\vert\vec{p}\vert}(iD_{k})+\frac{p_{i}}{2\vert\vec{p}\vert}\bigg ) c (\vec{p},t)+\nonumber\\
&+2x^{0}\int d^{3}\vec{p}\,\overline{c(\vec{p},t)}\bigg(\vert \vec{p}\vert (i D_{i}) -\frac{ip_{i}}{2\vert\vec{p}\vert}\bigg)c(\vec{p},t)+\nonumber\\
&+(x^{0})^{2}\int d^{3}\vec{p}p_{i}\vert c (\vec{p},t)\vert^{2}
\end {align}
On-shell form of eqs. (\ref{al26}) reads
\begin {align} 
\label {al27}
G=\int d^{3}\vec{p}\,\overline{c(\vec{p})}\hat{G}c(\vec{p})
\end {align}
with $\hat {G} $ being one of the operators (\ref{al12}), (\ref{al16}) with $\lambda=-\frac{1}{2}$.
\section{Semiclassical approximation} 
\label{IV}
\par Assuming, in the spirit of semiclassical approach, that $\vert c(\vec{p},t)\vert$ is strongly peaked at some $\vec{p}_{c}(t)$ and putting
\begin {align} 
\label {al28}
c(\vec{p},t)=\vert c(\vec{p},t)\vert e^{-i\omega(\vec{p},t)}
\end {align}
one finds \cite{b11}, \cite{b19}:
\begin {align} 
\label {al29}
S&=\int d t \int d^{3}\vec{x}\psi^{+}(\vec{x},t)\bigg(i\frac{\partial}{\partial t}-i\sigma_{k}\frac{\partial}{\partial x^{k}}\bigg)\psi (\vec{x},t)\equiv\nonumber\\
&\equiv \int d t \mathcal{L} (t)
\end {align}
\begin {align} 
\label {al30}
\mathcal{L} (t)\equiv \vec{p}_{c}\!\cdot\! \dot{\vec{x}}_{c}-\vert\vec{p}_{c}\vert-\vec{\alpha}(\vec{p}_{c})\!\cdot\!\dot{\vec{p}}_{c}
\end {align}
with
\begin {align} 
\label {al31}
\vec{\alpha}(\vec{p})\equiv-iu_{+}^{+}(p)\vec{\nabla}_{p}u_{+}(p)
\end {align}
\begin {align} 
\label {al32}
\vec{x}_{c}\equiv iu^{+}_{+}(p_{c})\vec{\nabla}_{p_{c}}u_{+}(p_{c})+\vec{\nabla}_{p_{c}}\omega (\vec{p}_{c},t)
\end {align}
\par The next step is to derive the semiclassical form of eqs. (\ref{al26}). To this end we follow the method used in Ref. \cite{b19} to obtain the semiclassical action (\ref{al29})-(\ref{al32}). We omit here the details; however, some points should be mentioned. We are dealing with semiclassical approximation and not the classical limit sketched in Sec.~\ref{II}. After restoring explicit $\hslash$ dependence we neglect all terms proportional to $\hslash$ except those hidden in the definitions of $\vec{x}_{c}$, eq. (\ref{al32}), or entering the value of helicity. Taking this into account we arrive at the following formulae
\begin {align} 
\label {al33}
\vec{\mathcal{P}}=&\vec{p}_{c}\nonumber\\
\mathcal{P}_{0}=&\vert\vec{p}_{c}\vert\nonumber\\
\mathcal{M}_{ij}=&p_{ci}x_{c}^{\phantom {c} j}-p_{cj}x_{c}^{\phantom {c}i}+\epsilon_{ijk}\frac{p_{ck}}{2\vert\vec{p}\vert}\nonumber\\
\mathcal{M}_{0i}=&\vert\vec{p}_{c}\vert x_{c}^{\phantom{c}i}+x^{0}p_{ci}\nonumber\\
\mathcal{D}=&-p_{ck}x_{c}^{\phantom {c} k}-x^{0}\vert \vec{p}_{c}\vert\nonumber\\
\mathcal{K}_{0}=&-\vert \vec{p}_{c}\vert x_{c}^{\phantom {c}k}x_{c}^{\phantom {c}k} + 2x^{0}p_{c}^{\phantom {c}k}x_{c}^{\phantom {c}k}-(x^{0})^{2}\vert \vec{p}_{c}\vert\nonumber\\
\mathcal{K}_{i}=&-p_{ci}x_{c}^{\phantom {c} k}x_{c}^{\phantom {c} k}+2x_{c}^{\phantom {c}i}x_{c}^{\phantom {c}k}p_{ck}-\epsilon _{ikl}\frac{x_{c}^{\phantom {c} k}p_{cl}}{\vert\vec{p}_{c} \vert}+\nonumber\\
&+2x^{0}\vert\vec{p}_{c}\vert x_{c}^{\phantom {c}i}+(x^{0})^{2}p_{ci}
\end {align}
\par In order to find the symmetry transformations generated by the integrals (\ref{al33}) let us remind that given the Hamiltonian form of the action
\begin {align} 
\label {al34}
S=\int \big(-\Theta_{\alpha}(\xi)d\xi_{\alpha}-H(\xi)dt\big ),
\end {align}
the symplectic structure is defined by the matrix
\begin {align} 
\label {al35}
\omega_{\alpha\beta}\equiv \partial_{\beta}\Theta_{\alpha}-\partial_{\alpha}\Theta_{\beta}
\end {align}
Our semiclassical $S$ is defined by eqs. (\ref{al29}), (\ref{al30}). The resulting Poisson brackets are given by eqs. (6) and (7) with $\lambda=-\frac{1}{2}$; this conclusion agrees with the form of quantum momentum and position operators (cf. eqs. (\ref{al9}) and (\ref{al11})). Once the Poisson structure is determined it is straightforward to find the symmetry transformations. The action of Poincare algebra coincides with that found in Refs. \cite{b11}-\cite{b20}. For dilatations we obtain
\begin {align} 
\label {al36}
&\delta\vec{x}_{c}=\delta \rho\bigg(\vec{x}_{c}-\frac{x^{0}\vec{p}_{c}}{\vert \vec{p}_{c}\vert}\bigg)\nonumber\\
&\delta \vec{p}_{c}=-\delta\rho\cdot \vec{p}_{c}
\end {align}
where $\delta\rho$ is an infinitesimal parameter. 
\par Special conformal transformations look slightly more complicated. They read
\begin {align} 
\label {al37}
&\delta \vec{x}=-\delta c^{0}\bigg (\frac{(\vec{x}^{2}) \vec{p}}{\vert \vec{p}\vert}+ \frac{\vec{x}\times\vec{p}}{\vert\vec{p}\vert^{2}}-2x^{0}\vec{x}+(x^{0})^{2}\frac{\vec{p}}{\vert \vec{p}\vert}\bigg)\nonumber\\
&\delta\vec{p}=\delta c^{0}(2\vert\vec{p}\vert \vec{x}-2x^{0}\vec{p})
\end {align}
for $\mathcal{K}_{0}$ and
\begin {align} 
\label {al38}
\delta \vec{x}=&\delta \vec{c} (\vec{x}^{2})+(\delta \vec{c}\cdot \vec{p})\bigg (\vec{x}\times \frac{\vec{p}}{\vert \vec{p}\vert^{3}}\bigg)-(\delta \vec{c}\times\vec{p})\frac{(\vec{x}\cdot \vec{p})}{\vert \vec{p}\vert^{3}}+\nonumber\\
&-2 (\delta \vec{c}\cdot \vec{x})\cdot \vec{x}+\frac{\delta \vec{c}\times \vec{x}}{\vert \vec{p}\vert}-\bigg(\delta \vec{c}\cdot\frac{(\vec{x}\times \vec{p})}{\vert \vec{p}\vert^{3}}\bigg) \vec{p}+\nonumber\\
&+2x^{0}(\delta \vec{c}\cdot \vec{x})\frac{\vec{p}}{\vert \vec{p}\vert}+2x^{0}\frac{(\delta \vec{c}\times \vec{p})}{\vert \vec{p}\vert^{2}}-(x^{0})^{2}\frac{\vec{p}}{\vert \vec{p}\vert}\nonumber\\
\delta \vec{p}=&-2(\delta\vec{c}\cdot \vec{p}) \vec{x}+2 \delta \vec{c}(\vec{x}\cdot \vec{p})+2 (\delta \vec{c}\cdot \vec{x})\vec{p}+\nonumber\\
&+\frac{\delta \vec{c}\times \vec{p}}{\vert \vec{p}\vert}-2x^{0}\vert \vec{p}\vert \delta \vec{c}
\end {align}
for $\mathcal{\vec{K}}$. In eqs. (\ref{al37}) and (\ref{al38}) we skipped for simplicity the subscript $"c"$; also all vectors have upper indices. 
\par It is straightforward to check that eqs. (\ref{al37}), (\ref{al38}) provide on-shell symmetries of semiclassical action.
\section{Conclusions} 
\label{V}
\par We have considered here the case of free fermions. The extension to the fermions interacting with external electromagnetic field is rather straightforward; however, it has some interesting aspects related to gauge invariance and will be considered in subsequent paper. It is also possible, although rather troublesome, to derive the semiclassical transformation rules directly from those of covariant wave functions following the reasoning presented in Ref. \cite{b20}.
\par In order to check the consistency of eqs. (\ref{al36})-(\ref{al38}) it is sufficient to verify that the semiclassical generators (\ref{al33}) obey, up to higher orders in $\hslash$, the (Poisson) commutation rules for conformal algebra. This is quite simple because one can compare the generators (\ref{al33}) with the classical expressions (\ref{al15}) and note that any power of $\lambda$ is accompanied by the same power of $\hslash$.
\par As it has been already noticed the extension to the case of higher helicity fermions in external field is problematic because the relevant (free) wave equations are of order $2\vert\lambda\vert $. One may attempt to rewrite them as a system of first order ones but this calls for introducing auxiliary variables which are subsequently eliminated by the relevant constraints. The problem with such an approach is that it is again not clear how the electromagnetic field could be switched on without destroying these constraints.
\\
\\
{\bf Acknowledgments}
\\
The research has been supported by the grant 2016/23/B/ST2/00727 of National Science Center, Poland.\\
We are grateful to Professors Krzysztof Andrzejewski and Cezary Gonera for helpful and interesting discussions.

\end{document}